\documentclass{aa}
\usepackage{graphicx}
\begin{document}
\title{Red Giant Branch Stars as Probes of Stellar Populations I: 2MASS
Calibration and Application to 2MASS GC01}

   \subtitle{}

   \author{Valentin~D.~Ivanov\inst{1}
          \and
          Jordanka~Borissova\inst{2}
          }

   \offprints{V.~D.~Ivanov}

   \institute{European Southern Observatory,
              Ave. Alonso de Cordova 3107, Casilla 19, Santiago 19001, Chile\\
              \email{vivanov@eso.org}
         \and
              Institute of Astronomy, Bulgarian Academy of Sciences, and
              Isaac Newton Institute of Chile, Bulgarian Branch,
              72~Tsarigradsko Chauss\`ee, BG\,--\,1784 Sofia, Bulgaria\\
              \email{jura@haemimont.bg}
             }

   \date{Received ...; accepted ...}

   \authorrunning{Ivanov \& Borissova}
   \titlerunning{2MASS Calibration of the RGB Parameters}

\abstract{
The near-infrared behavior of the red giant branch (RGB hereafter)
as a function of abundance is examined with an unprecedented large
sample of 27 Galactic globular clusters with Two Micron All Sky
Survey photometry. We propose a new simplified analyses, involving
the zero point of the RGB slope fit, and derive calibrations for the
RGB slope, zero point, and tip. The weak metallicity sensitivity of
the zero point leads to a ``fan''-like diagram for obtaining the
abundance distributions in resolved stellar systems,
and reddening estimates. Finally, we apply the new calibrations to
the recently discovered Galactic globular cluster 2MASS GC01, to
derive [Fe/H]$_{H96}=-1.19\pm0.38$ mag. The uncertainty is dominated
by the severe foreground contamination. We estimate an extinction of
$A_V=21.07\pm2.20$ mag toward the cluster.

}

\maketitle

\section{Introduction}

The red giant branch (RGB hereafter) stars are among the brightest red
stars in stellar systems, older than a few Gyrs. These stars appear in
almost all galaxies, including II~Zw~40 (\"{O}stlin \cite{ostlin}),
considered until recently as the best candidate for a primeval galaxy.
Therefore the red giants are a promising tool for probing the parameters
of old populations and the history of star formation in any galaxy.

Galactic globular clusters, with their single age and metallicity, are
the ideal sites for calibrating the RGB parameters. Since Da Costa \&
Armandroff (\cite{daC90}) provided the first reliable calibration of
the position of the RGB versus metallicity, there has been a significant
advancement, both because of the improvement of the astronomical
instrumentation and the development of the corresponding theory.

The infrared waveband is particularly compelling for such studies, in
comparison with the optical one, because of the potential to probe the
stellar populations of systems with high foreground and/or intrinsic
extinction. The
relatively small size and field of view of the IR arrays have made it
more difficult to carry out photometry of large areas, and to compile
uniform samples, necessary to calibrate the RGB parameters reliably, in
comparison with the optical region. The Two Micron All Sky
Survey\footnote{See http://www.ipac.caltech.edu/2mass/releases/second/
doc/explsup.html} (2MASS hereafter) offered for the first time such an
opportunity.

Previous calibrations suffered a number of drawbacks. Kuchinski, Frogel,
\& Terndrup (\cite{kft95}) and Kuchinski \& Frogel (\cite{kf95}) studied
only metal rich globular clusters, with [Fe/H]$\geq-$1. Later on, Ivanov
et al. (\cite{iva00}) added to their sample three metal poor globular
clusters. Ferraro et al. (\cite{ferr00}) used exceptional quality data
but in a photometric system, based on unpublished standards by Ian Glass
(South African Astronomical Observatory) with no available
transformations to any of the other systems. The similarity of their
final results to those of Ivanov et al., who used the CIT system, leads
to a conclusion that the two photometric systems are not radically
different.

To expand the basis of the RGB studies, we used high quality uniform
photometry of Galactic globular clusters from the 2MASS point source
catalog, and assembled a sample of RGBs of clusters with well known
distances and reddening. We calibrated the behavior of the RGB slope,
zero point, and tip with metallicity in the 2MASS photometric system. For
the first time we offer such calibration in a well defined photometric
system with all-sky coverage. We present a ``fan'' diagram, suitable for
abundance distribution analysis. Our results complement the recent work
of Cho \& Lee (\cite{cho01}) and Grocholski \& Sarajedini (\cite{gro01}).
who explored the properties of the RGB bump.

\section{2MASS Calibration of the RGB Parameters}

\subsection{Sample\label{SecSample}}

Infrared photometry of about 80 of the 147 globular cluster, listed in
Harris (\cite{harr96}, revision June 22, 1999) is currently available from
the First and the Second 2MASS incremental releases. However, many of the
observed clusters are rendered unsuitable for this project because they
suffer from one or more of the following drawbacks:
\begin{enumerate}
\item Severe foreground contamination. This refers mostly to globular
clusters in the general direction of the Galactic center.
\item The photometry of some distant halo clusters or highly reddened
clusters is not sufficiently deep.
\item Compact globular clusters have unresolved cores, limiting the
number statistics, and leading to crowding problems.
\end{enumerate}

After a visual inspection of the luminosity functions and color-magnitude
diagrams (CMD hereafter) of all available globular clusters, we selected
a subset of 27 objects (Table~\ref{TblSample}) for which reliable
estimates of at least one RGB parameters could be obtained. To calculate
the absolute magnitudes, and the intrinsic colors of stars, we adopted
the reddening and distance estimates from the compilation of Harris
(\cite{harr96}), and the reddening law of Rieke \& Lebofsky (\cite{rl85};
used throughout this paper).

\begin{table}[t]
\begin{center}
\caption{Globular cluster sample. The Columns give: cluster ID, color
excess E$_{B-V}$, distance modulus, tidal radius R$_t$ (all from the
compilation of Harris \cite{harr96}), and abundances in various scales
(see the references).}
\label{TblSample}
\tabcolsep=5pt
\begin{tabular}{c@{}c@{}ccc@{}c@{}r}
\hline
\multicolumn{1}{c}{NGC} &
\multicolumn{1}{c}{E$_{B-V}$} &
\multicolumn{1}{c}{(m-M)$_v$} &
\multicolumn{1}{c}{[Fe/H]} &
\multicolumn{1}{c}{[Fe/H]} &
\multicolumn{1}{c}{[M/H]} &
\multicolumn{1}{c}{R$_{t}$} \\
\multicolumn{1}{c}{ID} &
\multicolumn{1}{c}{mag} &
\multicolumn{1}{c}{mag} &
\multicolumn{1}{c}{H96} &
\multicolumn{1}{c}{CG97} &
\multicolumn{1}{c}{F99} &
\multicolumn{1}{c}{$\arcmin$} \\
\hline
 104 &0.04&13.37&$-$0.76& $-$0.70 & $-$0.59 &47.25\\
 288 &0.03&14.69&$-$1.24& $-$1.07 & $-$0.85 &12.94\\
 1851&0.02&15.47&$-$1.22& $-$1.08 & $-$0.88 &11.70\\
 1904&0.01&15.59&$-$1.57& $-$1.37 & $-$1.22 & 8.34\\
 2298&0.14&15.59&$-$1.85& $-$1.74 &($-$1.54)& 6.48\\
 5024&0.02&16.38&$-$1.99&($-$1.82)&($-$1.62)&21.75\\
 5139&0.12&13.97&$-$1.62&($-$1.38)&($-$1.18)&44.85\\
 5466&0.00&16.15&$-$2.22&($-$2.14)&($-$1.94)&34.24\\
 6121&0.36&12.83&$-$1.20& $-$1.19 & $-$0.94 &32.49\\
 6144&0.32&16.06&$-$1.73& $-$1.49 &($-$1.29)&33.25\\
 6171&0.33&15.06&$-$1.04& $-$0.87 & $-$0.70 &17.44\\
 6205&0.02&14.48&$-$1.54& $-$1.39 & $-$1.18 &25.18\\
 6273&0.37&15.85&$-$1.68&($-$1.45)&($-$1.25)&14.50\\
 6284&0.28&16.70&$-$1.32&($-$1.10)&($-$0.90)&23.08\\
 6356&0.28&16.77&$-$0.50&($-$0.66)&($-$0.50)& 8.97\\
 6441&0.44&16.62&$-$0.53&($-$0.67)&($-$0.51)& 8.00\\
 6624&0.28&15.37&$-$0.42&($-$0.64)&($-$0.48)&20.55\\
 6637&0.16&15.16&$-$0.71& $-$0.68 & $-$0.55 & 8.35\\
 6656&0.34&13.60&$-$1.64&($-$1.41)&($-$1.21)&29.97\\
 6715&0.14&17.61&$-$1.59&($-$1.35)&($-$1.15)& 7.47\\
 6779&0.20&15.65&$-$1.94&($-$1.75)&($-$1.55)& 8.56\\
 6809&0.07&13.87&$-$1.81& $-$1.61 & $-$1.41 &16.28\\
 6838&0.25&13.75&$-$0.73& $-$0.70 & $-$0.49 & 9.96\\
 6864&0.16&16.87&$-$1.32&($-$1.10)&($-$0.90)& 7.28\\
 7078&0.10&15.37&$-$2.25& $-$2.12 & $-$1.91 &21.50\\
 7089&0.06&15.49&$-$1.62&($-$1.38)&($-$1.18)&21.45\\
 7099&0.03&14.62&$-$2.12& $-$1.91 & $-$1.71 &18.34\\
\hline
\multicolumn{7}{l}{References for [Fe/H]:}\\
\multicolumn{7}{l}{H96 - Harris (\cite{harr96}; Zinn scale);}\\
\multicolumn{7}{l}{CG97 - Carretta \& Gratton (\cite{cg97});}\\
\multicolumn{7}{l}{F99 - Ferraro et al. (\cite{ferr99}) }\\
\hline
\end{tabular}
\end{center}
\end{table}

A special care was taken to assure the uniformity of the metallicity data.
Harris (\cite{harr96}) based his compilation on the system established by
Zinn \& Wsst (\cite{zw84}), Zinn (\cite{zinn85}), and Armandroff \& Zinn
(\cite{az88}). Carretta \& Gratton (\cite{cg97}) suggested that this scale
may overestimate the abundance of metal-rich globular clusters, and
developed a new one. We list their measurements in Table~\ref{TblSample} as
well. Values in brackets are estimates based on the transformation they
derive, to the Zinn \& West scale (see Eq. 7 of Carretta \& Gratton).


Ferraro et al. (\cite{ferr99}) argued that the total amount of heavy
elements is a better metallicity measurement, because it accounts
naturally for the opacity variations, which depend on the total
metallicity, not just the iron abundance (Salaris, Chieffi, \& Straniero
\cite{sal93}). They developed a news scale, designed to measure the that.
For the clusters, absent in their sample, we adopted their prescription
for $\alpha$-element enhancement (Sect. 3.4 in Ferraro et al.; estimates
bracketed in Table~\ref{TblSample}).

Since the metallicities originate from different sources, and the
corresponding uncertainties are often not quoted, we adopted an uniform
error of 0.20 dex for all measurements. This is a typical value of the
accuracy of the abundance estimates.

\begin{figure}
\resizebox{\hsize}{!}{\includegraphics{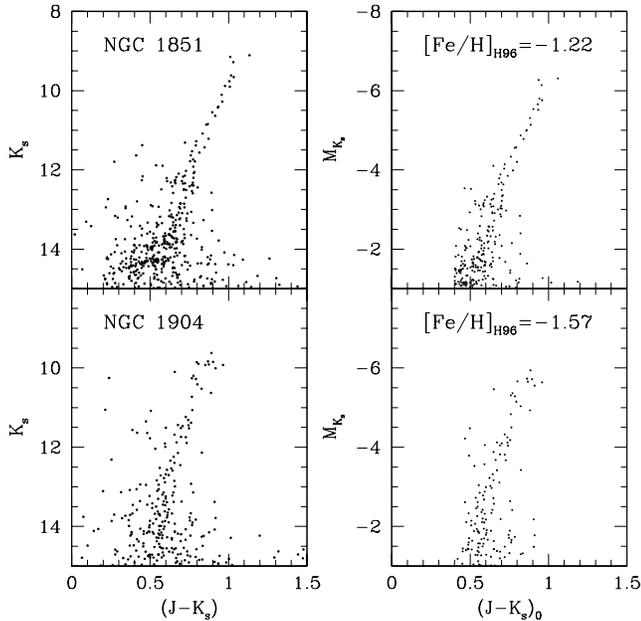}}
\caption{Color-magnitude diagrams for a subset of metal poor globular
clusters. The left panels show the raw data. The right panels show the
reddening corrected CMDs, converted to absolute magnitudes. The foreground
contamination and stars with $(J-K_S)_0<0.4$ mag have been removed.}
\label{FigCMDs1}
\end{figure}

\begin{figure}
\resizebox{\hsize}{!}{\includegraphics{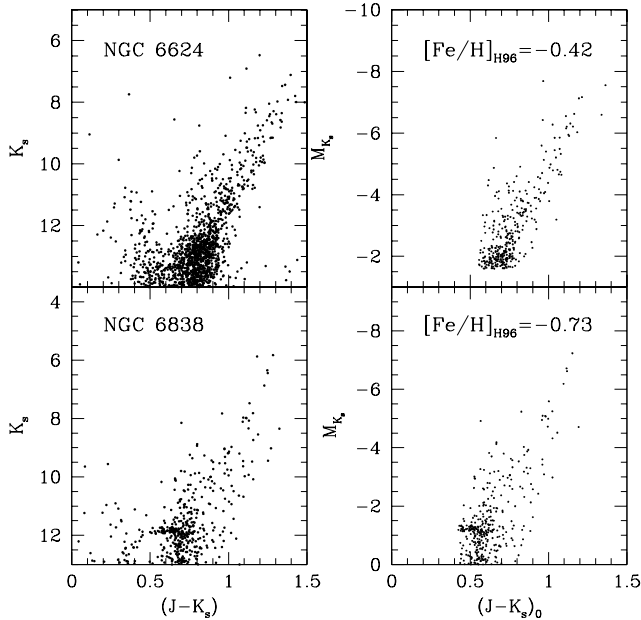}}
\caption{Same as Figure~\ref{FigCMDs1} for metal rich clusters.}
\label{FigCMDs2}
\end{figure}

The first step before estimating the RGB parameters was to eliminate
statistically the foreground contamination. As many stars were removed
randomly from the CMD of the globular clusters, as the number of stars,
present in a nearby field with the same area as the globular cluster
field. The statistical removal is not reliable in case of severely
contaminated globular clusters, where the number of foreground stars was
comparable to the number of cluster members. A subset of CMDs for four
clusters is shown in Figure~\ref{FigCMDs1} and \ref{FigCMDs2}. In
addition, we removed the stars within 15 arcsec from the clusters
centers, to minimize the crowding effects.

\subsection{RGB Slope and Zero Point\label{SecSlope}}

Da Costa \& Armandroff (\cite{daC90}) prepared a set of standard RGBs on
the V versus V-I CMD, for metallicity estimates of globular clusters in
the optical. The technique has been expanded toward the near infrared
wavebands, where the giant branch is linear, and therefore, more obvious
and easier to define compared with the optical or optical-infrared CMDs.
The slope is related to the effective temperature of the stars along the
RGB, and $\rm T_{eff}$ in turn depends on the opacity, and the heavy
element abundance. The slope is a reddening and distance-free metallicity
indicator. It is significantly less demanding in terms of observing time
and telescope collecting area than the spectroscopic methods, but
unlike them it can be applied reliably only to uniform groups of stars.

We followed the procedure to determine the RGB slope, described in
Ivanov et al. (\cite{iva00}), fitting the RGB on the $[K_s,(J-K_s)]$ CMD
with a linear equation: $J-K_s=a\times K_s+b$. Only the stars above the
horizontal branch were included. We applied a least square method,
taking into account the uncertainties along both axis. Two iterations
were carried, and all the stars outside $5\sigma$ from the first fit
were removed from the calculations.

For the purpose of fitting the RGB of an individual cluster, the zero
point is equivalent to previous calibrations of the RGB colors at
fixed $K$-band magnitude levels (e.g. Ivanov et al. \cite{iva00}; Ferraro
et al. \cite{ferr00}), although more straightforward. To derive the
zero-point relation to the abundance, we used only the clusters which
suffer minimal extinction $({\rm E}_{B-V}\leq$\,0.3 mag).

Notably, the fit to the zero point shows a very small variation with
metallicity: 0.11 mag for [Fe/H]$_{H96}$ varying from -2 to -0.5. For
comparison, $(J-K_S)_0$ at $K_S=-5.5$ mag varies by 0.32 mag for the
same metallicity range (Ferraro et al. \cite{ferr00}). This result
offers the possibility for a more reliable reddening estimates than in
case of RGB color calibrations.

\begin{figure}
\resizebox{\hsize}{!}{\includegraphics{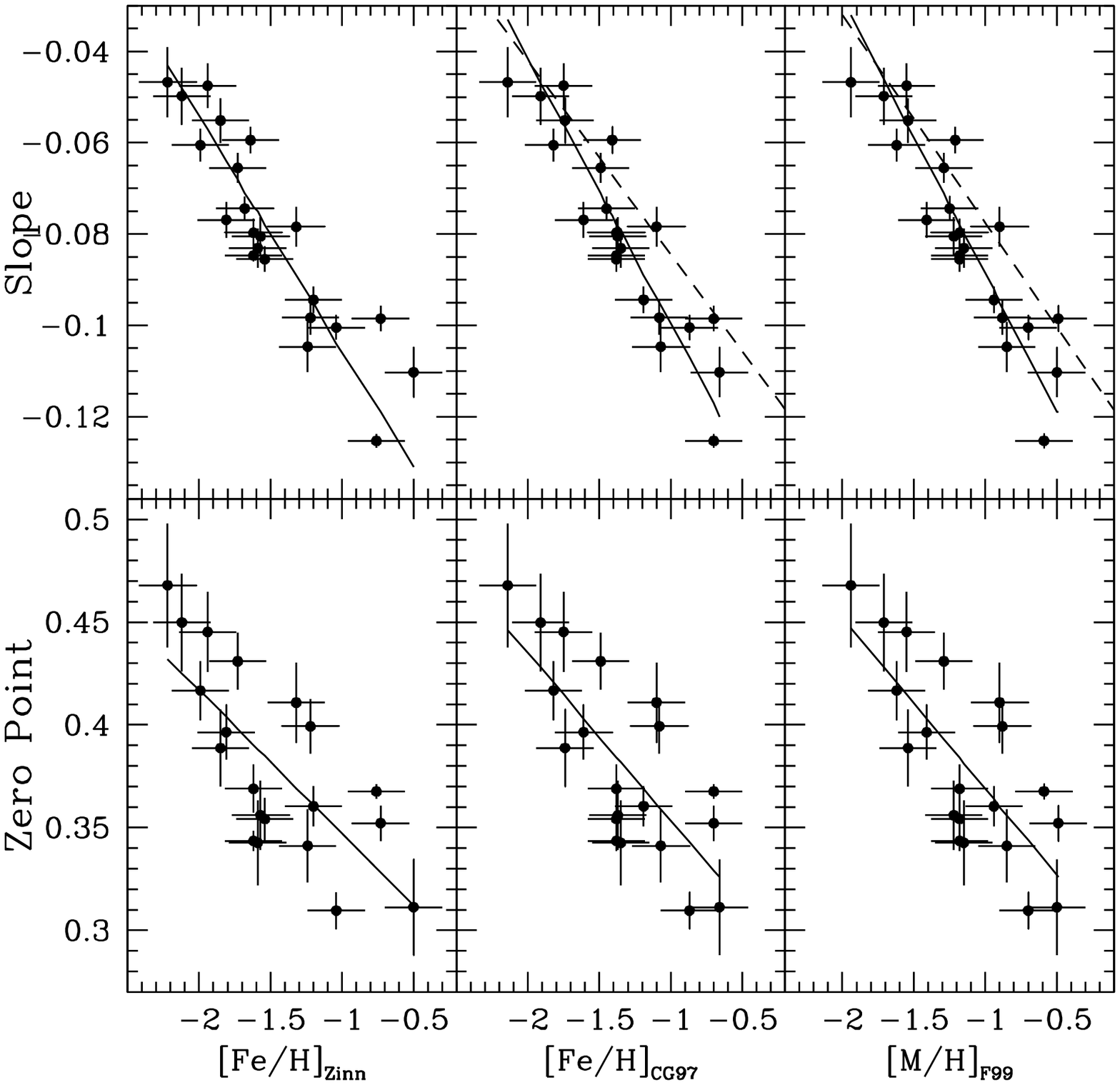}}
\caption{Relation of the [Fe/H] in various metallicity scales
versus RGB slope on $[M_{K_s},(J-K_s)_0]$ diagram. The bars indicate
$1\sigma$ uncertainties. The dashed line is the fit of Ferraro et al.
(\cite{ferr00}) drawn without transformation of colors. The zero point
errorbars show only the statistical errors from the fits to the RGB,
omitting the 0.2 mag added in quadrature to account for the uncertain
distance moduli and reddening corrections.}
\label{FigSlopeZP}
\end{figure}

The behavior with metallicity of the RGB slope and zero point are
demonstrated in Figure~\ref{FigSlopeZP}. Fitting coefficients for
individual cluster RGBs are listed in Table~\ref{TblSlope}.

\begin{table}[t]
\begin{center}
\caption{Linear fits to the RGB in the form of: $J-K_s=a\times K_s+b$.
The radii R$_{max}$ of the region around the cluster center used
in the fit, and the root mean square of the fit are given.}
\label{TblSlope}
\begin{tabular}{c@{}c@{}r@{ }r@{ }r}
\hline
\multicolumn{1}{c}{NGC} &
\multicolumn{1}{c}{R$_{max}$} &
\multicolumn{1}{c}{a} &
\multicolumn{1}{c}{b} &
\multicolumn{1}{c}{r.m.s.} \\
\multicolumn{1}{c}{ID} &
\multicolumn{1}{c}{$\arcmin$} &
\multicolumn{1}{c}{Slope} &
\multicolumn{1}{c}{Zero Point} &
\multicolumn{1}{c}{mag} \\
\hline
104  & 10 & $-$0.125(0.002) & 0.368(0.004) & 0.054 \\
288  &  7 & $-$0.105(0.006) & 0.341(0.018) & 0.048 \\
1851 &  5 & $-$0.098(0.004) & 0.399(0.013) & 0.095 \\
1904 &  7 & $-$0.081(0.004) & 0.356(0.017) & 0.088 \\
2298 &  6 & $-$0.055(0.005) & 0.389(0.019) & 0.104 \\
5024 & 11 & $-$0.061(0.004) & 0.417(0.014) & 0.107 \\
5139 &  7 & $-$0.085(0.001) & 0.344(0.005) & 0.063 \\
5466 & 10 & $-$0.047(0.008) & 0.468(0.030) & 0.082 \\
6121 & 10 & $-$0.094(0.003) & 0.360(0.010) & 0.091 \\
6144 &  4 & $-$0.066(0.003) & 0.431(0.014) & 0.088 \\
6171 &  8 & $-$0.101(0.003) & 0.310(0.009) & 0.096 \\
6205 & 10 & $-$0.086(0.003) & 0.354(0.010) & 0.068 \\
6273 &  3 & $-$0.074(0.003) &              & 0.099 \\
6356 &  2 & $-$0.110(0.005) & 0.311(0.023) & 0.109 \\
6656 &  3 & $-$0.059(0.003) &              & 0.074 \\
6715 &  3 & $-$0.083(0.004) & 0.343(0.021) & 0.143 \\
6779 &  3 & $-$0.048(0.005) & 0.445(0.020) & 0.069 \\
6809 & 10 & $-$0.077(0.004) & 0.396(0.013) & 0.064 \\
6838 &  3 & $-$0.099(0.003) & 0.352(0.009) & 0.085 \\
6864 &  3 & $-$0.078(0.004) & 0.411(0.020) & 0.110 \\
7089 &  9 & $-$0.080(0.003) & 0.369(0.012) & 0.090 \\
7099 &  8 & $-$0.050(0.006) & 0.450(0.024) & 0.047 \\
\hline
\multicolumn{5}{l}{Note: Fitting errors are given in brackets. The Zero}\\
\multicolumn{5}{l}{Points for NGC\,6273 and 6656 are omitted because}\\
\multicolumn{5}{l}{of the large reddening toward this cluster.}\\
\hline
\end{tabular}
\end{center}
\end{table}

\subsection{RGB Tip}

The tip of the RGB is a well known distance indicator. Mould \& Kristian
(\cite{mou86}), Lee (\cite{l93}) and Lee et al. \cite{lfm93} pioneered the
method in the optical, measuring the corresponding jump in the luminosity
function in external galaxies. Later it was applied to number of resolved
stellar systems, among the most challenging of which were some Virgo cluster
members (Harris et al. \cite{harr98}) and NGC~3379 (Sakai et al.
\cite{sak97}).

This is a statistically demanding technique, requiring 50-100 stars per
bin.  The limited size of infrared arrays explains the difficulty to
apply it in the near infrared. Although it does cover a large area, the
2MASS photometry can not alleviate the intrinsic problem of the small
number of giants in globular clusters. Thus, we assume that the
brightest cluster member represents the RGB tip.

In most of the cases the tip is obvious, but sometimes additional
criteria had to be applied to determine the brightest stars. We took
advantage of the linearity of the RGB in the near infrared, and excluded
from the considerations bright stars that deviated from the color of the
RGB at a given magnitude level, predicted by the RGB slope fit by more
than 0.5 mag. We also excluded some extremely bright stars, with
luminosity higher than the rest of the RGB by 2-3 mag. They were obvious
foreground contamination. Finally, the red variables from Clement et al.
(\cite{cle01}) were excluded .

The formal uncertainties of the stellar magnitudes given in 2MASS were
discarded, since they do not represent well the uncertainty in the tip
magnitude. Instead, we adopted the difference in the magnitudes of the
two brightest stars, accounting for the possibility that the brightest
star may be a non-member. This led to typical error values of 0.2-0.4 mag,
much larger than the 2MASS errors.

\begin{figure}
\resizebox{\hsize}{!}{\includegraphics{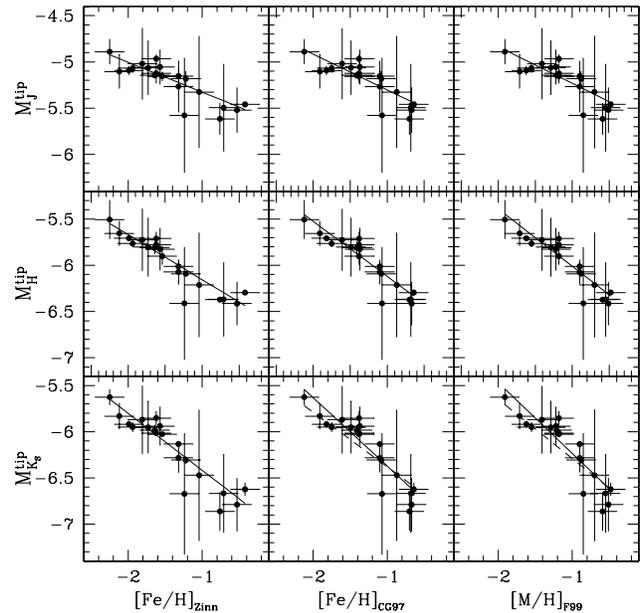}}
\caption{Relation of the [Fe/H] in various metallicity scales versus
RGB tip absolute magnitude. The bars indicate $1\sigma$ uncertainties,
with the Y-axis errors including the uncertainties due to the
reddening and distance. The dashed line is the fit of Ferraro et al.
(\cite{ferr00}) drawn without transformation of colors.}
\label{FigTip}
\end{figure}

The RGB tip magnitudes for 20 globular clusters are given in
Table~\ref{TblTip}. The behavior of the RGB tip with metal abundance
is shown in Figure~\ref{FigTip}.

\begin{table}[t]
\begin{center}
\caption{Estimated RGB tip absolute magnitudes.}
\label{TblTip}
\begin{tabular}{c@{ }c@{ }c@{ }c}
\hline
\multicolumn{1}{c}{NGC} &
\multicolumn{1}{c}{$J$} &
\multicolumn{1}{c}{$H$} &
\multicolumn{1}{c}{$K_s$} \\
\hline
104  & $-$5.616(0.171)& $-$6.372(0.017)& $-$6.862(0.209)\\
288  & $-$5.578(0.621)& $-$6.412(0.610)& $-$6.672(0.656)\\
1851 & $-$5.182(0.083)& $-$6.092(0.024)& $-$6.306(0.040)\\
1904 & $-$5.055(0.189)& $-$5.826(0.231)& $-$5.940(0.211)\\
5024 & $-$5.093(0.049)& $-$5.708(0.006)& $-$5.919(0.035)\\
5139 & $-$5.124(0.106)& $-$5.778(0.040)& $-$6.015(0.003)\\
6144 & $-$5.062(0.294)& $-$5.803(0.318)& $-$5.953(0.280)\\
6171 & $-$5.326(0.605)& $-$6.213(0.565)& $-$6.471(0.711)\\
6205 & $-$5.156(0.050)& $-$5.901(0.092)& $-$6.027(0.020)\\
6284 & $-$5.267(0.277)& $-$6.073(0.133)& $-$6.283(0.156)\\
6441 & $-$5.521(0.246)& $-$6.413(0.235)& $-$6.788(0.288)\\
6624 & $-$5.459(0.019)& $-$6.295(0.027)& $-$6.623(0.072)\\
6637 & $-$5.496(0.474)& $-$6.369(0.401)& $-$6.667(0.426)\\
6656 & $-$5.146(0.023)& $-$5.809(0.067)& $-$5.984(0.059)\\
6779 & $-$5.067(0.050)& $-$5.765(0.036)& $-$5.950(0.054)\\
6809 & $-$5.020(0.385)& $-$5.725(0.335)& $-$5.870(0.367)\\
6864 & $-$5.154(0.072)& $-$6.012(0.047)& $-$6.133(0.007)\\
7078 & $-$4.890(0.136)& $-$5.506(0.211)& $-$5.626(0.087)\\
7089 & $-$4.966(0.046)& $-$5.711(0.072)& $-$5.853(0.076)\\
7099 & $-$5.105(0.183)& $-$5.655(0.134)& $-$5.832(0.142)\\
\hline
\multicolumn{4}{l}{Note: The difference between the two brightest}\\
\multicolumn{4}{l}{stars is given in brackets.}\\
\hline
\end{tabular}
\end{center}
\end{table}

\subsection{Results}

\begin{figure}
\resizebox{\hsize}{!}{\includegraphics{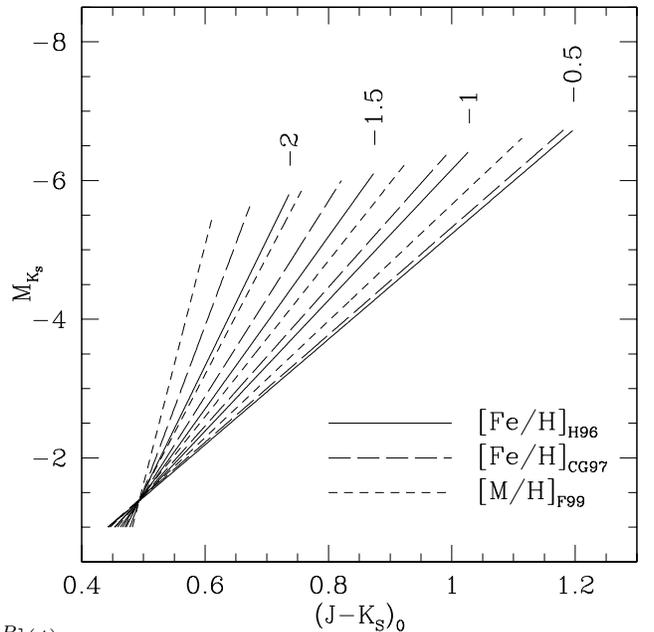}}
\caption{``Fan Diagram'' - a grid of RGBs on the reddening and
distance corrected color-magnitude diagram, for the three
metallicity scales, as indicated.}
\label{FigGrid}
\end{figure}

The behavior of the derived RGB parameters with metallicity was fitted
with linear equations, taking into account the errors along both axes.
The coefficients are given in Table~\ref{TblCoeff}. Figure~\ref{FigGrid}
shows a ``fan''-like grid of RGBs in a reddening and distance corrected
CMD, for the three metallicity scales discussed in Section~\ref{SecSample}.
It demonstrates that the RGB behavior can be reduced to a simple rotation
around a nearly-fixed point. This comes to no surprise, since the zero
point of the RGB fit is almost independent of the abundance
(Section~\ref{SecSlope}).

The grid allows to obtain the metallicities of individual sdtars in
resolved systems in the infrared, and to obtain the metallicity
distributions as it has been done before in the optical (e.g. Saviane et
al. \cite{sav00}). The linear representation of the RGB in absolute
$K_S$ magnitude versus the intrinsic $(J-K_S)_0$ color is:

\begin{equation}
M_{K_S} = RGB_{ZP} + RGB_{Sl} \times (J-K_S)_0
\end{equation}

Both the slope $RGB_{Sl}$ and the zero point $RGB_{ZP}$ were calibrated
as functions of the abundance:

\begin{equation}
RGB_{Sl} = a_0^{Sl} + a_1^{Sl} \times {\rm [Fe/H]}
\end{equation}
\begin{equation}
RGB_{ZP} = a_0^{ZP} + a_1^{ZP} \times {\rm [Fe/H]}
\end{equation}

After substituting Eq. 2 and 3 in 1, and solving for the stellar
metallicity:

\begin{equation}
{\rm [Fe/H]} = [ M_{K_S} - a_0^{Sl}\times (J-K_S)_0 - a_0^{ZP} ] ~/~
   [ a_1^{Sl} \times (J-K_S)_0 + a_1^{ZP} ]
\end{equation}

This equation should be used with caution. Small photometric errors are
crucial for accurate [Fe/H] estimates. The upper part of the infrared CMD
is better suited for determining individual metallicities because near the
root of the RGB the abundance relation degenerates. The exact faint limit
is determined by the quality of the photometry, and by the presence of
the horizontal branch at $M_{K_S}\sim -1.5$ mag.

\begin{table}[t]
\begin{center}
\caption{Polynomial fits to the metallicity behavior of the RGB slope,
zero point, bump, and tip: Y=a$_0$+a$_1$$\times$X.}
\label{TblCoeff}
\begin{tabular}{l@{  }l@{  }c@{ }c@{ }c@{ }c}
\hline
\multicolumn{2}{c}{Variables} &
\multicolumn{2}{c}{Coefficients} &
\multicolumn{1}{c}{r.m.s.} &
\multicolumn{1}{c}{$\rm N_{pts}$} \\
\multicolumn{1}{c}{X} &
\multicolumn{1}{c}{Y} &
\multicolumn{1}{c}{a$_0$($\sigma_{{\rm a}_0}$)} &
\multicolumn{1}{c}{a$_1$($\sigma_{{\rm a}_1}$)} &
\multicolumn{2}{c}{} \\
\hline
 H66&$\rm RGB_{Sl}$&$-$0.157(0.009)&$-$0.051(0.006)&0.002&22\\
CG97&$\rm RGB_{Sl}$&$-$0.158(0.010)&$-$0.058(0.007)&0.002&22\\
 F99&$\rm RGB_{Sl}$&$-$0.149(0.009)&$-$0.060(0.008)&0.002&22\\
 H66&$\rm RGB_{ZP}$&$+$0.277(0.152)&$-$0.070(0.104)&0.007&20\\
CG97&$\rm RGB_{ZP}$&$+$0.272(0.155)&$-$0.082(0.120)&0.007&20\\
 F99&$\rm RGB_{ZP}$&$+$0.285(0.137)&$-$0.084(0.122)&0.007&20\\
 H66& $\rm J_{tip}$&$-$5.650(0.187)&$-$0.323(0.121)&0.025&20\\
CG97& $\rm J_{tip}$&$-$5.690(0.210)&$-$0.387(0.151)&0.026&20\\
 F99& $\rm J_{tip}$&$-$5.631(0.191)&$-$0.399(0.158)&0.026&20\\
 H66& $\rm H_{tip}$&$-$6.641(0.186)&$-$0.486(0.121)&0.025&20\\
CG97& $\rm H_{tip}$&$-$6.712(0.210)&$-$0.594(0.153)&0.024&20\\
 F99& $\rm H_{tip}$&$-$6.631(0.193)&$-$0.620(0.162)&0.024&20\\
 H66& $\rm K_{tip}$&$-$7.032(0.212)&$-$0.615(0.134)&0.032&20\\
CG97& $\rm K_{tip}$&$-$7.109(0.243)&$-$0.739(0.171)&0.034&20\\
 F99& $\rm K_{tip}$&$-$7.003(0.224)&$-$0.768(0.181)&0.034&20\\
\hline
\multicolumn{5}{l}{Note: Errors are given in brackets. Metallicity}\\
\multicolumn{5}{l}{notation is the same as in Table~\ref{TblSample}.}\\
\hline
\end{tabular}
\end{center}
\end{table}

\section{2MASS GC01\label{SecGC01}}

Hurt et al. (\cite{hur99}, \cite{hur00}) reported the serendipitous
discovery of 2MASS GC01 (hereafter GC01) in 2MASS data. It is a heavily
obscured globular cluster, laying in the Milky Way disk, in the general
direction of the Galactic center. Ivanov, Borissova, \& Vanzi
(\cite{iva00b}) determined $\rm (m-M)_0=12.4-14.0,$ and \
${\rm A}_V=20.9-18.8,$ assuming $\rm [Fe/H]=-0.5$ and $-2.0,$
respectively. The main source of uncertainty in these estimates was
the unknown metallicity of the cluster, although the location of GC01
suggested that it might be a metal rich object.

The first step toward a metallicity estimate of GC01 was to remove the
foreground stars contamination. Unlike the clusters we used to derive the
RGB parameters calibrations, the contamination here is severe, reaching
$\sim30$\% in the RGB region. We performed 2000 foreground substractions,
and estimated the RGB slope for each realization separately. This method
yields distributions of the RGB slope and tip, and the widths of these
distributions measure the respective uncertainties.

To carry out this procedure we defined the CMD area, encompassing the
RGB: $8.0\leq K_S\leq14.0$ mag, and $2.5\leq J-K_S\leq6.0$ mag. Then we
divided it into 0.2 mag square bins. Experiments with different bin
sizes indicated that any value between 0.2 and 1.0 mag leads to the same
conclusions.

Next, we constructed CMDs for the cluster field, and for a surrounding
field with an equal area. To minimize the crowding effects we omitted
the stars within 15 arcsec from the cluster center. The outer limit of
the cluster field was constrained by the cluster diameter (3.3$\pm$0.2
arcmin, Hurt et al. \cite{hur99}, \cite{hur00}). We carried out our
calculations to two values of the outer radii: 1.0 and 1.5 arcmin.
Smaller values limit the number statistics, and larger ones increase the
fraction of the foreground contamination. The foreground field was
defined as a circular annulus with $10.49-10.54$ or $10.43-10.54$ arcmin
size, respectively for 1.0 and 1.5 armin cluster fields.

Finally we counted the stars in each bin, and subtracted randomly from
the ``cluster'' bins as many stars as were present in the ``field'' bins.
If the latter bin had more stars than the former one, we subtracted stars
from the nearby bins, again in a random way.

\begin{figure}
\resizebox{\hsize}{!}{\includegraphics{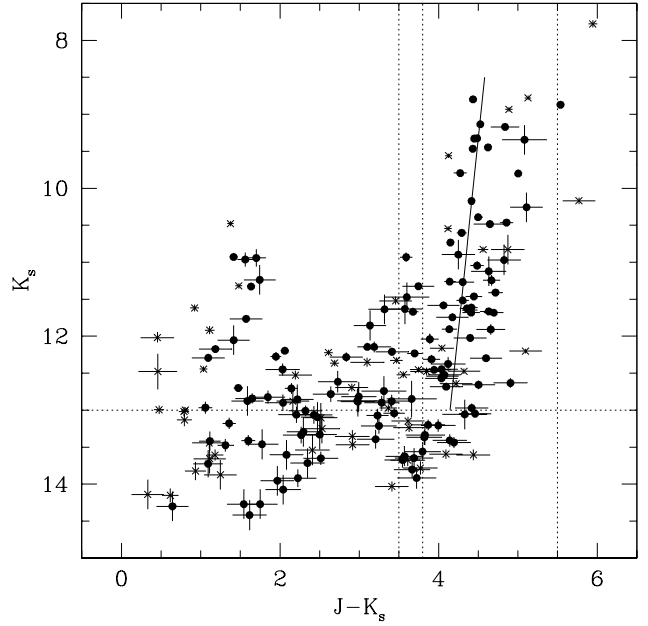}}
\caption{2MASS color-magnitude diagram of GC01. Solid dots are stars
from 0.25 to 1 arcmin from the cluster center - cluster+foreground.
X's indicate the stars within 10.49 to 10.54 arcmin from the center -
pure foreground. $1\sigma$ errors are indicated. The solid line
represents the average RGB fit, described in Section~\ref{SecGC01}.
Clearly, it is dominated by the stars with minimal uncertainties.
Dotted lines are the adopted limits.}
\label{FigGC01CMD}
\end{figure}

We carried out a linear fit on the RGB stars in the foreground subtracted
CMD in the same manner as for the calibration clusters (see
Section~\ref{SecSlope}). For stars without error measurements in 2MASS
we adopted $\sigma=0.20$ mag. Using the linearity of the RGB, we imposed a
faint limit of the stars, included in the fit, just above the horizontal
branch level. The luminosity function of GC01 (Ivanov et al. \cite{iva00b},
Figure 5) indicates that the horizontal branch is at $K_S$=13.0-13.2 mag.
To minimize the uncertainties of the RGB slope, we also imposed color
limits on the stars we used in the fit. The red one was set to
$J-K_S$=5.5 mag, and has no effect on the slopes because of the
negligible number of stars to the red of the RGB. The results are
somewhat more sensitive to the blue limit. We choose to impose
$J-K_S$=3.50 and 3.80 mag, because of the well-defined limit of the RGB
at this colors (Figure~\ref{FigGC01CMD}).

A summary of the results for the RGB slope and zero point GC01 is
presented in Figure~\ref{FigRGBpars}. The RGB tip is omitted because
it is sensitive only to the adopted cluster radius. Clearly, the
effects from the assumed parameters are smaller or comparable with
the uncertainties originating due to the foreground contamination.
We calculated the error-weighted averages for the realizations with
$K_S^{lim}=12.5-13.0,$ to avoid the possible influence of the cluster
horizontal brunch. The determined RGB parameters for GC01 are:

\begin{figure}
\resizebox{\hsize}{!}{\includegraphics{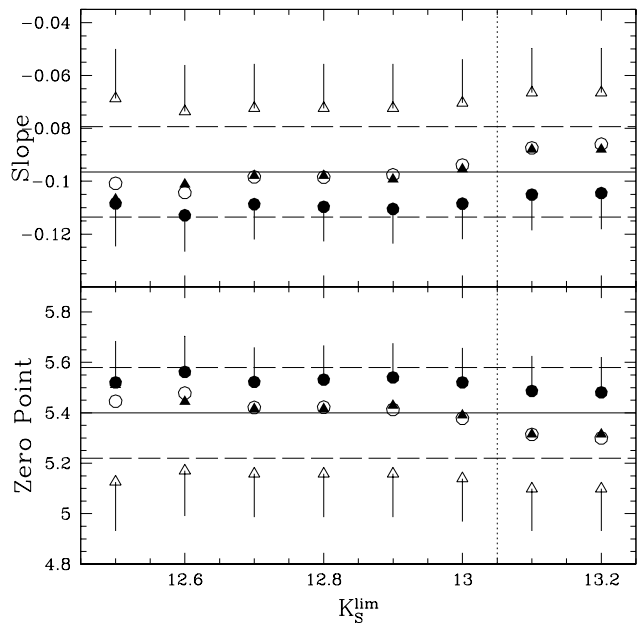}}
\caption{Behavior of the RGB slope (top) and zero point (bottom)
for different parameters of the foreground subtraction and fits. The
horizontal axis is the lower limit of the stars used in the RGB fit.
Triangles indicate cluster radius of 1.0 arcmin, and circles indicate
1.5 arcmin. Open symbols indicate a blue limit of $J-K_S$=3.25 mag,
and solid symbols of $J-K_S$=3.00 mag. Only some $1\sigma$ errors
of individual points are shown to avoid crowding. The solid lines
indicate the weighted averages, and the dashed lines indicate their
$1\sigma$ errors. The values to the right of the vertical dotted
line were discarded because they were affected by the horizontal
branch.}
\label{FigRGBpars}
\end{figure}

\begin{enumerate}
\item $< {\rm Slope} >\,= -0.0965\pm0.0171$
\item $< {\rm Zero Point} >\, = 5.40\pm0.18$
\item $< {\rm Tip} >\,= 8.87\pm0.13$
\end{enumerate}

Using the equations, given in Table~\ref{TblCoeff}, we obtain:
[Fe/H]$_{H96}$= $-1.19\pm$0.38, [Fe/H]$_{CG97}$= $-1.06\pm$0.34, and
[M/H]$_{F99}$= $-0.88\pm$0.32. The errors given here include both
the uncertainties in the slope, and in the calibrations. GC01 appears
to resemble closely NGC~104.

The estimated absolute magnitude for the RGB tip, for a cluster with
such abundance is $M_{K_S}= -6.30\pm0.35$ mag, using our new
calibration for the Zinn metallicity scale. This leads to a distance
modulus of $(m-M)_0+A_{K_S}=15.17\pm0.38$ mag consistent with Ivanov
et al. (\cite{iva00b}). We refrain from further considerations based
on the RGB tip because of the poor statistics at the brighter end of
the RGB. Instead, we will adopt $(m-M)_0+A_{K_S}=15.0\pm0.4$ mag, a
result of interpolation between the values for [Fe/H]=$-1.0$ and
$-2.0,$ in Table 2 of Ivanov et al. (\cite{iva00b}).

We can also verify if the cluster reddening is consistent with the
previous estimates. First, we subtract the RGB slope equations, written
for the GC01 in apparent, and in absolute magnitudes. Respectively:
\begin{equation}
(m_J-m_{K_S}) = {\rm RGB_{ZP}^{app}} + {\rm RGB_{Sl}} \times m_{K_S}
\end{equation}
\begin{equation}
(M_J-M_{K_S})_0 = {\rm RGB_{ZP}^{abs}} + {\rm RGB_{Sl}} \times M_{K_S}
\end{equation}
Here the notation is the same as in Table~\ref{TblCoeff}, and
${\rm RGB_{ZP}^{abs}}=0.36\pm0.20$ mag, for the [Fe/H]$_{H96}$ given
above. The result of the subtraction is:
\begin{equation}
E(J-K_S) = {\rm RGB_{ZP}^{app}} - {\rm RGB_{ZP}^{abs}} + {\rm RGB_{Sl}}
\times [(m-M)_0+A_{K_S}]
\end{equation}
Substituting, we obtain: $E(J-K_S) = 3.58\pm0.37$ mag, close to the
values of 3.49 mag, interpolated as above from Ivanov et al.
(\cite{iva00b}). Our new estimate corresponds to $A_V=21.07\pm2.20$
mag, and $A_{K_S}=2.36\pm0.25$ mag.

\section{Summary}

The behavior of the RGB in the infrared was quantified based on an
unprecedented large sample of 2MASS photometry of Milky Way globular
clusters. The RGBs were fitted by straight lines. We produced new
calibrations of the RGB slope, tip, and - for the first time - zero
point, as functions of abundance. The introduction of the zero point
streamlines greatly the RGB analyses in comparison with the traditional
approach where RGB colors at given levels were used. Notably, the zero
point is fairly insensitive to the abundance, varying by only 0.11 mag
over a range from [Fe/H]$_{H96}=-2$ to $-0.5.$ We present a ``fan''-like
diagram, suitable for analyses of the metallicity spread in resolved
stellar systems.

The derived calibrations were applied to estimate the metal abundance
of the recently discovered globular cluster GC01. It is a particularly
challenging object because of the severe foreground contamination. We
removed it with a random procedure, and used the RGB slope of the
remaining pure cluster population to derive
[Fe/H]$_{H96}$=$-1.19\pm$0.38. The uncertainty is dominated by the
foreground contamination, and albeit large, allows to exclude extremely
high abundances, expected from the cluster location. GC01 is likely to
be moderately metal poor. The RGB tip and zero point yield distance
modulus and extinction, consistent with our previous estimates.

\begin{acknowledgements}
This publication makes use of data products from the Two Micron All Sky
Survey, which is a joint project of the University of Massachusetts and
the Infrared Processing and Analysis Center, funded by the National
Aeronautics and Space Administration and the National Science Foundation.
The authors thank Dr. Ivo Saviane for the useful discussions, and the
referee Dr. M. G. Lee for the comments that halped to improve the paper.
\end{acknowledgements}

\end{document}